# Closed-loop Error Correction Learning Accelerates Experimental Discovery of Thermoelectric Materials

Hitarth Choubisa[1,†], Md Azimul Haque,[2,†], Tong Zhu[1], Lewei Zeng[1], Maral Vafaie[1], Derya Baran[2,*], Edward H Sargent[1,*]

[1] *Department of Electrical and Computer Engineering University of Toronto, Toronto, Ontario, Canada*

[2] *King Abdullah University of Science and Technology (KAUST), Physical Science and Engineering Division, KAUST Solar Center (KSC), Thuwal, Saudi Arabia*

* Correspondence: Derya Baran (derya.baran@kaust.edu.sa) and Edward Sargent (ted.sargent@utoronto.ca)

† H.C. and M.A.H. contributed equally to this paper.

**Abstract**:

The exploration of thermoelectric materials is challenging considering the large materials space, combined with added exponential degrees of freedom coming from doping and the diversity of synthetic pathways. Here we seek to incorporate historical data and update and refine it using experimental feedback by employing error-correction learning (ECL). We thus learn from prior datasets and then adapt the model to differences in synthesis and characterization that are otherwise difficult to parameterize. We then apply this strategy to discovering thermoelectric materials where we prioritize synthesis at temperatures < 300°C. We document a previously unreported chemical family of thermoelectric materials, PbSe:SnSb, finding that the best candidate in this chemical family, 2 wt% SnSb doped PbSe, exhibits a power factor more than 2x that of PbSe. Our investigations show that our closed-loop experimentation strategy reduces the required number of experiments to find an optimized material by as much as 3x compared to high-throughput searches powered by state-of-the-art machine learning models. We also observe that this improvement is dependent on the accuracy of prior in a manner that exhibits diminishing returns, and after a certain accuracy is reached, it is factors associated with experimental pathways that dictate the trends.

# Main:

Thermoelectric materials convert thermal energy to electricity based on the Seebeck effect, in which a thermal gradient leads to electric current generation[1]. Their capacity to convert waste heat into electricity provides a route to recover the energy lost in mechanical and electrical processes[2]. Today's best-performing thermoelectric materials are particularly effective at high temperatures, whereas further progress is necessary to enable energy harvesting in low-temperature consumer applications such as IoT devices and wearables[3,4].

The design and discovery of thermoelectric materials is challenging in view of the large chemical space[5,6], non-convex composition-property mapping[7] and nonlinear effects of dopants on material properties[8,9]. Large-scale chemical space searches for thermoelectric materials have been pursued in the past using high-throughput Density Functional Theory (DFT) based simulations[10–16] and machine learning (ML) based property prediction searches[8,17–19]. Relatively few efforts have combined theory-obtained and experimentally measured datasets.

It will be a priority to learn from combined theory-experiment data sets to account for differences in synthesis conditions that can lead to different morphologies – morphologies well-known to influence thermoelectric performance. Valuable and important precedents exist for refining models using closed-loop strategies[20–24]. However, these studies either require pre-defined functional forms for the data fusion methods to integrate available data and acquired data; or the presence of relevant sampled data for estimating the property distribution. The first factor limits how information can be transferred between the two data sources whereas the latter can be computationally expensive[20] or impossible to curate from available datasets. Consequently, accurate learning of the material-property relationship is limited, which prevents efficient material exploration quantified by accuracy scores on data acquired and ranked according to performance.

We hypothesize that a data-driven strategy for information transfer between the available and acquired data can overcome the first limitation. The problem of uniformly sampling the material space to have an informed prior can then be addressed independently and becomes equivalent to the development of accurate ML surrogate models. We combine these and demonstrate a two-step error correction learning approach: learning from existing datasets and iterative refinements with new experimental results (**Figure 1** for details on the discovery process).

We apply this strategy to the discovery of earth-abundant and low-temperature thermoelectric materials (300K) through optimization of the experimental power factors. We explore the material space through the synthesis of new inorganic compounds, doping and alloying leading to the discovery of a new family of low-temperature thermoelectric materials PbSe:SnSb, with the predicted best candidate in this chemical family exhibiting a power factor x2 that of PbSe and largest among any previously reported low-temperature ($< 300°C$) synthesized materials. In the process, we find that chemical representation as well as the accuracy of ML models on available datasets play quantitatively significant roles in accelerating closed-loop materials discovery. Exploration using DFT of the origins of higher power factor shows the interaction of Se and Sb in PbSe:SnSb leading to a reduced hole effective mass and larger power factors. The work shows how ECL based closed loop approaches can be used to account for factors that are traditionally difficult to parameterize in materials discovery.

**Error-correction learning with experiments:**

In a standard ML-based materials screening pipeline, one learns the mapping between given structures/compositions and the outcome variable(s) of interest[25–28]. The trained model is used to screen large chemical spaces and rank the promising candidates. The top-ranked candidates are then validated experimentally[8,29]. However, synthesis conditions such as the synthesis method and morphologies differ from one experimental laboratory to another. These practical yet unavoidable factors render predictions of ML models inaccurate. As a result, more experiments are necessary to explore the chemical space and find the best experimentally validated material candidates.

With this in mind, we formulate the above as an error-correction problem: the observed outcomes of our in-lab experiments y for two given compositions $c_1$ and $c_2$ can be expressed as:

$$y = \begin{cases} 0 & if \quad PF(c_1, T_1) \geq PF(c_2, T_2) \\ 1 & if \quad PF(c_1, T_1) < PF(c_2, T_2) \end{cases} \quad (1)$$

where $c_1$ and $c_2$ are the materials we want to compare, $T_1$ and $T_2$ are the temperatures we want to make the comparison at and $PF(\cdot)$ represents the in-lab measured power factor (or generally any other property of interest). $y$ is dependent on the synthesis parameters and methods. Hence, it is not easy to merge it with existing datasets coherently and train the ML model on the combined data. Instead, we propose to model the observed outcome $y$ using an error-correction strategy,

$$y \sim g_\theta(X_1, X_2, T_1, T_2, m) \qquad (2)$$

where $X_1$ and $X_2$ are vectorial representations for the two material compositions $c_1$ and $c_2$ we want to compare, $T_1$ and $T_2$ are the temperatures we want to make the comparison at, $PF(\cdot)$ represents the power factor (or any other property of interest) and $m$ is the *prior model* trained on existing datasets as one would do in a high-throughput ML search pipeline. The exact form of error-correction function is unknown. To avoid parametric biases, we use a dense neural network $g_\theta$ that can learn from data to model the error-correction function (**Figure 1a**). The proportion of error correction that is achievable depends on the quantity of data, model $m$ and methods used for the vectorial representation of chemical compositions.

**Discovery of large power factor thermoelectrics:**

We use an experimental dataset of thermoelectrics[30] compiled by a recent study[8] to train the ML-based data generator. The dataset reports thermoelectric metrics such as conductivity ($\sigma$) and PF for various doped and alloyed materials. We compare multiple chemical representations for training the ML model and choose the one with the best cross-validation score as the prior model for driving closed-loop experiments. We test three different chemical representations – Magpie[31], Roost[28] and CrabNet[27]. **Figure 1b** shows a high-level summary of how these three representations differ. We use transfer learning for training with Roost and CrabNet approaches due to the relatively smaller size of the dataset. The features extracted after the global pooling layer are used to represent different compositions and as input features to train the model (refer to **SI notes 1-2** for details on ML model training and the models considered).

Interestingly, the *prior model* trained using Magpie showed the best cross-validation performance outperforming Roost and CrabNet (refer to **Figure 2a** for a comparison of various representations and ML models we trained and compared). We explore this observation by analyzing the distribution of the generated representations by each of the methods (SI **Figure S1**). Due to its min-max operators, we observe that Magpie generates a less smooth representation that can better capture the doping and alloying effects. On the other hand, Roost and CrabNet have smoother changes in the representation vectors for different materials and that is reflected through lower accuracies for power factor prediction, a property sensitive to small amounts of alloying and doping.

After training the *prior model*, we explore the chemical space to discover low-temperature large power factor materials. In our exploration, we limit ourselves to specific precursors and elements. We focused on materials synthesized below <300°C and avoided regulated materials from the list {As, Cd, Hg, Te, Tl}. Imposing these constraints limits the favorable chemical space. Chalcogenides present as part of the material precursors are further screened based on their power factors (PFs) reported in the Materials Project[32]. This results in the list of available precursors (**SI note 3** for the complete list).

As the first step, we measure the experimental PF for all the chalcogenide compositions. The error-correcting dense neural network (DNN) is then trained to error-correct the predictions across different chemical compositions and temperatures. We optimize the hyperparameters and estimate the generalization using 5-fold cross-validation accuracy. The most generalizable model is then used to explore a new set of compositions for peak performance at 300K that is fed back, and the cycle is repeated. We explored different material engineering strategies in every round: composites, alloys and dopants (**SI note 4** for more details). With every round of experimentation, the ability of the error-correction model to account for the experimental changes improves (**Figure 1c** for the schematic on workflow). Multiple rounds of the feedback loop and exploration finally yield (PbSe):(2 wt% SnSb) as the optimal composition (refer to the section below for details on experimental characterization). The results of hyperparameter optimizations are summarized in **Figure S2**.

We compare the effect of change in the *prior model* on the accuracy of the error-correction learning (ECL) model (**Figure 2b**). For this, we train the error-correction model on all the data collected from the experiments conducted before the last round of experiments. The accuracy is then evaluated on the last unseen round of experiments. We find that higher accuracy prior models lead to higher accuracy on error-corrected predictions. Interestingly, models sharing the same chemical representation also tend to perform similarly i.e., Magpie-based models, even though differing in prior model's accuracy show high error-corrected accuracies.

Similarly, both the CrabNet-based models show similar accuracies post error correction. This observation aligns with the discussion before on the origin of higher accuracy observed with Magpie based representation (**SI Figure 1)**. This indicates that choosing an appropriate representation is essential while driving closed-loop experimentation. We also compared these results to the case where we used a weight-sharing network instead of a DNN. The trends were similar but with lower accuracy (**Figure S3**).

Furthermore, the underperformance of the next best model CrabNet after error correction on the ranking of compounds indicates that at least one more round of experimentation would have been needed to reach the optimal candidate. (Figure 2b). This means that in this setting, the accurate prior model developed herein reduces the number of experiments by at least (1/5×100%)=20% (the total number of rounds of experiments in such a scenario would be 5). Finally, the best candidate composition $(PbSe)_{0.98}(SnSb)_{0.02}$ is ranked at 98$^{th}$ if the material candidates are ranked according to the order of decreasing power factors using the most accurate *prior model*. Thus, compared to the conventional high-throughput screening approach of testing the high-ranked predictions from the prior model, the current approach reduces the number of candidates that need to be tested by as much as 83% i.e., less than 1/3rd the number of experiments that would need to be conducted. We also compare our approach to a modified version of the previously proposed data fusion approach[20] in the **Table 1** (refer to **SI note 5** for implementation details).

| Method | Accuracy on the final-round data at 219°C |
|---|---|
| Probabilistic constraint data-fusion[20] | 0.57 |
| ECL (this study) | 0.99 |

**Table 1**: We tabulate the accuracy and compare a former approach that requires defining rules for data fusion to our approach that utilizes dense neural networks for error-correction.

**Experimental validation:**

While the synthesis of inorganic compounds by solution method is tedious due to the limited number of suitable metal salts and solubility issues, vacuum techniques require an extensive amount of time for optimization. As such, we adopted ball milling to synthesize some of the predicted compounds owing to its high yield and faster optimization rate. We tested the top candidates from ML predictions and evaluated the compounds' thermoelectric properties. The measured properties were fed to the error correction module to improve the accuracy.

We observe that highest electrical conductivity ($\sigma$) was obtained for SnSb with a low Seebeck coefficient (*S*). Stoichiometry change and forming composite with $PbI_2$ did not improve Seebeck of SnSb. However, its combination with PbSe resulted in balanced $\sigma$ and *S*. Temperature-dependent $\sigma$ and *S* (refer to the **characterization subsection of Methods** for more details) of the stoichiometric compounds, SnSe, $WSe_2$, GaTe and $CrZnTe_4$ exhibit semiconducting behavior while SnSb, PbSe, and $NbSe_2$ show metallic behavior. Notably, $CrZnS_4$ has extremely low electrical conductivity and consequently high Seebeck (**Figure 3a-c**). The highest electrical conductivity is observed in the case of SnSb (> 10000 S/cm) which leads to a very low Seebeck coefficient. PbSe exhibits the largest power factor among the stoichiometric compounds due to balanced electrical conductivity and Seebeck. Despite of extraordinary conductivity of SnSb, its modest power factor is a consequence of low Seebeck. In contrast, PbSe shows good Seebeck, but its electrical conductivity is not very high (We compare conductivities of many composites made and tested but not added to Figure 3b in **Figure S4)**.

Our experiments align with the prediction that small SnSb addition to PbSe can improve the power factor. We observe that small amounts of SnSb addition enhance the electrical conductivity of PbSe with a minor decrease in Seebeck (refer to **materials subsection** of Methods for details on synthesis and materials). Therefore, the PbSe/SnSb composition results in a high power factor of more than 850 µW/mK$^2$ at room temperature for 2 wt% SnSb, and good performance is observed in the whole temperature regime (**Figure 3d-f**). Such a high-power factor for the present composite is comparable to state-of-the-art polycrystalline thermoelectric materials prepared by similar methods (**Figure S5**). It should be noted that further performance improvement can be anticipated by additional processing techniques, such as spark plasma sintering, generally employed in the case of high-performance thermoelectric materials.

**Origins of high power factor:**

To understand the atomistic origin of the high performance of $(PbSe)_{0.98}(SnSb)_{0.02}$, we perform Density Functional Theory (DFT) calculations (refer to the **DFT subsection of Methods** for details). We obtain the unit cell for the PbSe supercell from Materials Project[32] and use it to construct a 216 atoms supercell. Experimental X-ray diffraction (XRD) indicates that a small amount of alloying does not change the structure (**Figure 4a**). We use this information to optimize the geometry of the SnSb doped PbSe supercell by relaxing only the atomic positions and keeping the lattice constants fixed using the Generalized Gradient Approximation (GGA) exchange-correlation (xc) functional (refer to **Figures S6-S7** for pictorial representation). We perform self-consistent HSE06 xc-functional calculations on the pristine and doped relaxed structures while incorporating spin-orbit coupling (SOC). Comparing the electronic structure of PbSe with PbSe:SnSb, we observe that the introduction of Sn and Sb modulates the density of states (DOS) near conduction band minimum (CBM) and valence band maximum (VBM) respectively. This modulation decreases the hole effective mass, improving the overall transport property of the PbSe:SnSb composite (**Figure 4b,c**) thereby resulting in a larger power factor. Please refer to **SI Figure S8-S9** for effective mass fitting plots.

## Conclusions:

In this study, we develop and demonstrate a two-step error correction learning approach to performing theory-driven closed-loop experimental exploration of thermoelectric materials. In the process, we train an ML model that outperforms existing models for accurate predictions of experimental power factors. We show that this improvement in accuracy leads to a significant reduction in the number of in-lab experiments. Our approach enables us to discover a novel thermoelectric material $(PbSe)_{0.98}:(SnSb)_{0.02}$ that exhibits a large power factor. DFT simulations show that the performance improvements are mainly a result of the chemical interactions and not structure distortions. While we demonstrate the efficacy of our proposed approach through the discovery of thermoelectric materials, it is equally applicable for the discovery of materials for other applications such as high conductivity and optoelectronics.

Figures

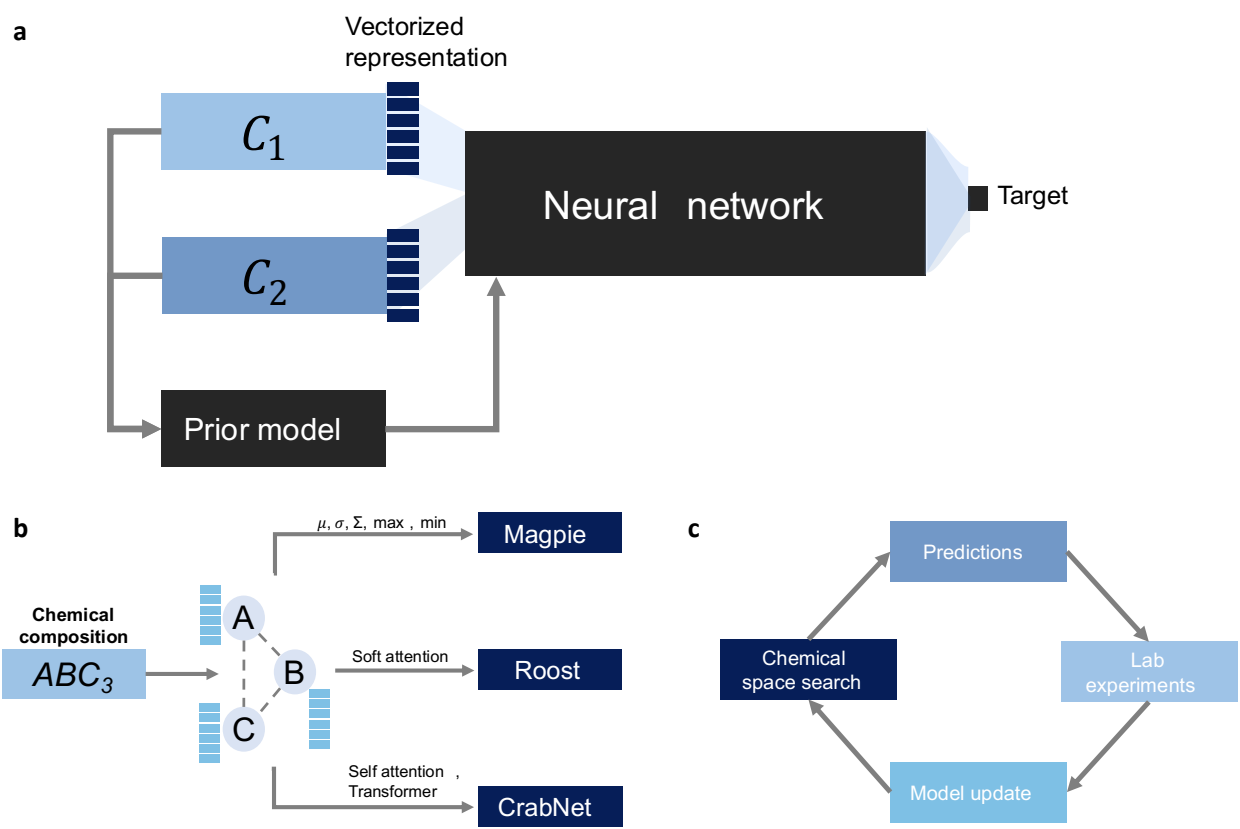

**Figure 1: Error correction learning to enable closed-loop materials discovery.** **(a)** shows our approach to perform error-correction learning (ECL) on predictions from a pre-trained machine learning model, addressed as *Prior model*. The error-correction network takes two chemical compositions ($C_1$ and $C_2$) as inputs and predicts if $C_1$ has a power factor larger than $C_2$. **(b)** describes the three approaches we used and compared for the purpose of vectorial representation of chemical compositions. Magpie[31] generates vectorial representation using basic statistical operations such summation, average, standard deviation, min and max operations on elemental properties. Roost[28] uses a soft-attention based mechanism and converts stoichiometric graphs to numerical vectorial representations. CrabNet[27] generates a numerical representation for chemical compositions using self-attention and transformer architecture. We benchmark all these approaches using their accuracy on literature data and accuracy on collected experimental data. **(c)** shows our closed loop approach: we start with the model and make predictions within the chemical space. The top candidates are used for lab experimentation and fed back to our error-correction module for improving the prediction accuracy.

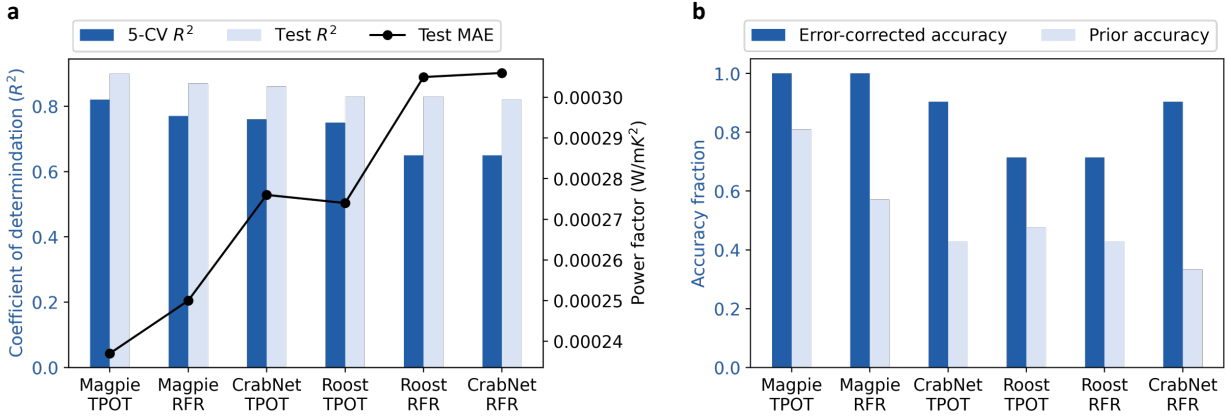

**Figure 2: Performance of machine learning models with and without ECL on available and acquired data (a)** shows the performance of the model with best set of hyperparameters chosen based on the cross-validation performance quantified using mean absolute error (MAE) and coefficient of determination ($R^2$). We found that Magpie[31] based featurization outperformed transfer learning using Roost[28] and CrabNet[27]. Here, the suffixes TPOT and RFR refer to the model used to belong to the one found by the TPOT library and Random Forest Regressor respectively. **(b)** compares the performance for each of model before and after performing ECL as described in the study. The prior accuracy is measured on across all compositions and temperatures based on power factor ranking. The error-corrected accuracy is reported based on ordering at 219°C. ECL improves the accuracy for all the cases. The improvement, however, depend upon both on featurization used and the initial prior model. The improvement varies as much as 0.16-0.40 (refer to **Figure S1** to analyze dependence on the neural network architecture).

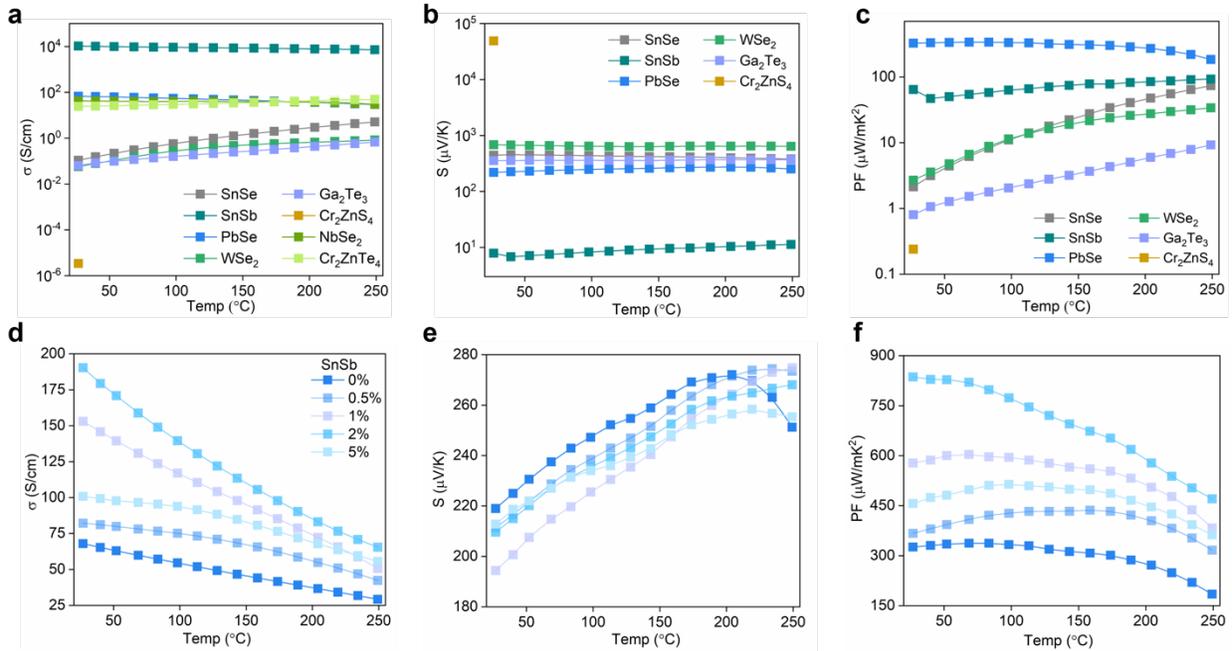

**Figure 3: Thermoelectric performance of materials measured experimentally (a, b, c)** show conductivity (σ), Seebeck coefficient (S), and Power Factor (PF) trends for stoichiometric compounds as a function of temperature. PbSe shows the highest power factor due to a balance between conductivity and Seebeck coefficient. **(d, e, f)** show $\sigma$, $S$ and PF trends for our predicted most promising compositions $(PbSe):(SnSb)_x$. We observe that the best composition has a power factor twice that of the undoped PbSe.

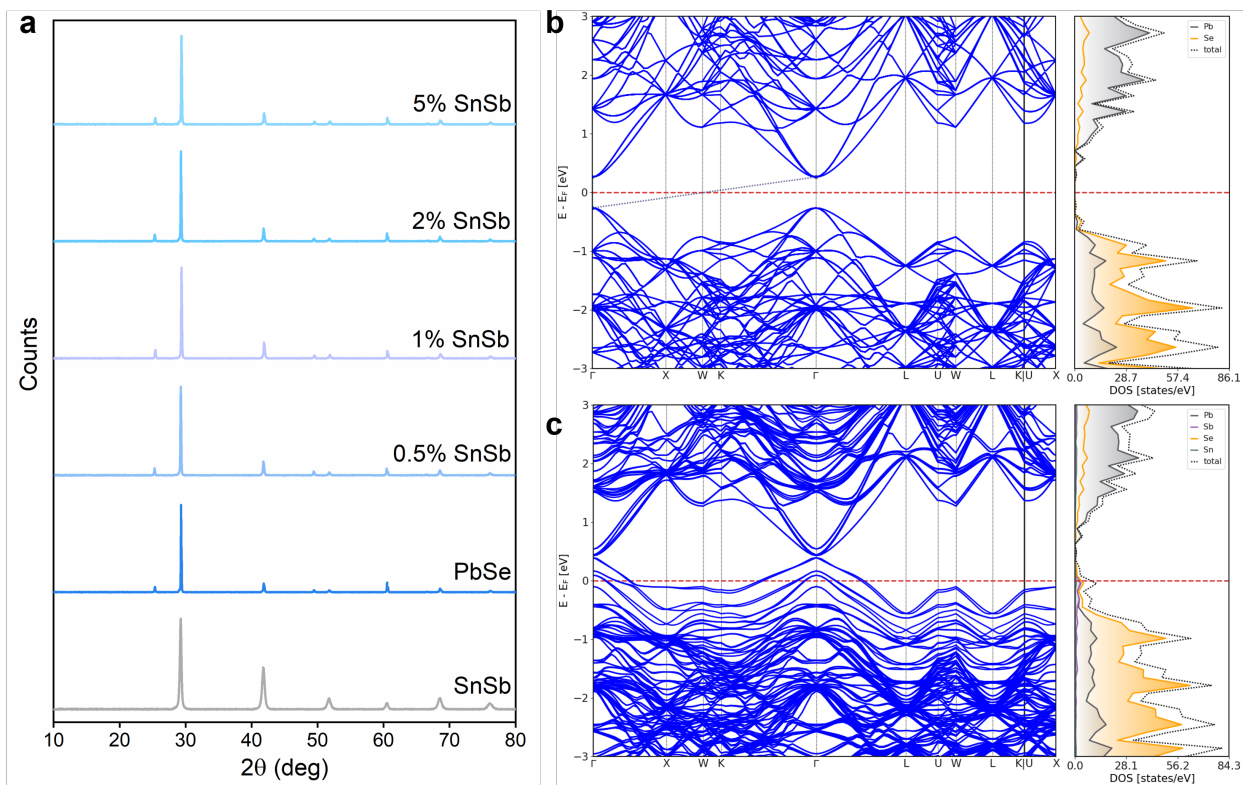

**Figure 4: Structural and DFT analysis of the best thermoelectric candidates (a)** shows XRD plots for (PbSe):(SnSb)$_x$ composite. **(b)** shows the band structure of PbSe along the high-symmetry path. **(c)** shows band structure of (PbSe):(2 wt% SnSb) along the high-symmetry path in the reciprocal space. The change in curvature of the state just below the defect state in the valence band leads to a reduction in the effective mass and therefore, improved transport properties (refer to **Figure S8-S9** for close-up).

## Methods:

### Data:

We use the dataset collected by DopNet[8] authors from the Materials Research Laboratory (MRL) for training the *prior models*. It is available at https://github.com/ngs00/DopNet and the original data can be found at http://www.mrl.ucsb.edu:8080/datamine/thermoelectric.jsp. Data based on our in-lab experiments is available at https://github.com/hitarth64/TherML.

### Code availability:

All the code and data used for this study are available as open source at https://github.com/hitarth64/TherML. We also provide all our trained models so that they can be used for future explorations.

### Materials:

Sn, Se, S, Zn, Te, and Cr powders were purchased from Sigma-Aldrich. PbSe, Sb, and $WSe_2$ were purchased from Alfa-Aesar. $NbSe_2$, GaSe, and $Ga_2Te_3$ were procured from American Elements. SnSe, SnSb, $Cr_2ZnS_4$, and $Cr_2ZnTe_4$ were synthesized by ball milling stoichiometric amounts of precursors for 8, 15, 32, and 32 hours at 30 Hz, respectively. All composites were obtained by ball milling the respective compounds (e.g. SnSe and SnSb) for 20 minutes at 30 Hz.

### Characterization:

Electrical conductivity and Seebeck measurements were performed on Netzsch SBA 548 Nemesis thermoelectric set up under He environment. Samples were prepared by loading the powders into a steel compaction die and compressing them to form pellets for thermoelectric measurements. All pellets were annealed at 250 °C inside an $N_2$ glovebox before thermoelectric measurements. The instrument uncertainty for electrical conductivity and Seebeck measurements are ±5% and ±7%, respectively. As a result of the very low electrical conductivity of $Cr_2ZnS_4$, room temperature Seebeck was measured using a manual setup consisting of Peltier devices and a thermocouple. For thermal voltage measurements, a homemade setup was used with Peltier devices and thermocouples to apply the temperature gradient across the sample, and the voltage was recorded using a Keithley 6517B electrometer. XRD was measured using Bruker D8 Advance.

**DFT:**

All DFT relaxations are performed using Vienna Ab initio Simulation Package (VASP) 6.2.1. We use Perdew-Burke-Ernzerhof (PBE)[33] exchange-correlation functional for geometry optimization of the material systems. The energy cut-off for the plane wave was set to 520 eV. The energy threshold for self-consistent energy convergence was set to $10^{-3}$ eV whereas the global convergence threshold was set to $10^{-2}$ eV energy difference between two successive ionic steps. A smearing width of 0.05 eV, consistent with the Materials Project, was chosen for calculations. Calculations were performed over a uniform k-points grid of $2 \times 2 \times 2$ generated using the Monkhorst scheme[34].

Given the optimized geometry, we performed static DFT calculation using hybrid HSE06[35] functional while incorporating the Spin-orbit coupling (SOC) effect over high-symmetry points of the reciprocal lattice using all-electron DFT code FHI-AIMS.

**Acknowledgment:**

This publication is supported by the King Abdullah University of Science and Technology (KAUST) Office of Sponsored Research (OSR) under Award No. OSR-CRG2018-3737. TOC was created by Ana Bigio, scientific illustrator at KAUST. ML models were trained using QUEST computing clusters located at Northwestern University. DFT calculations were performed both at the QUEST computing cluster located at Northwestern University and the Narval computing cluster which is part of Compute Canada and made accessible through University of Toronto. We thank Lydia Li for help in designing Figure 1.

**References:**

1. Jood, P., Ohta, M., Yamamoto, A. & Kanatzidis, M. G. Excessively Doped PbTe with Ge-Induced Nanostructures Enables High-Efficiency Thermoelectric Modules. *Joule* **2**, 1339–1355 (2018).

2. Tan, G., Ohta, M. & Kanatzidis, M. G. Thermoelectric power generation: from new materials to devices. *Philosophical Transactions of the Royal Society A* **377**, (2019).

3. Vostrikov, S., Somov, A. & Gotovtsev, P. Low temperature gradient thermoelectric generator: Modelling and experimental verification. *Appl Energy* **255**, 113786 (2019).

4. van Toan, N., Thi Kim Tuoi, T., van Hieu, N. & Ono, T. Thermoelectric generator with a high integration density for portable and wearable self-powered electronic devices. *Energy Convers Manag* **245**, 114571 (2021).

5. Davies, D. W. *et al.* Computational Screening of All Stoichiometric Inorganic Materials. *Chem* **1**, 617–627 (2016).

6. Choubisa, H. *et al.* Crystal Site Feature Embedding Enables Exploration of Large Chemical Spaces. *Matter* (2020) doi:10.1016/j.matt.2020.04.016.

7. Choubisa, H. *et al.* Accelerated chemical space search using a quantum-inspired cluster expansion approach. (2022) doi:10.48550/arxiv.2205.09007.

8. Na, G. S., Jang, S. & Chang, H. Predicting thermoelectric properties from chemical formula with explicitly identifying dopant effects. *npj Computational Materials 2021 7:1* **7**, 1–11 (2021).

9. Chakraborty, S. *et al.* Rational Design: A High-Throughput Computational Screening and Experimental Validation Methodology for Lead-Free and Emergent Hybrid Perovskites. *ACS Energy Lett* **2**, 837–845 (2017).

10. Ding, G., Gao, G. & Yao, K. High-efficient thermoelectric materials: The case of orthorhombic IV-VI compounds. *Scientific Reports 2015 5:1* **5**, 1–7 (2015).

11. Jain, A., Shin, Y. & Persson, K. A. Computational predictions of energy materials using density functional theory. *Nature Reviews Materials 2016 1:1* **1**, 1–13 (2016).

12. Zhu, H. *et al.* Computational and experimental investigation of TmAgTe2 and XYZ2 compounds, a new group of thermoelectric materials identified by first-principles high-throughput screening. *J Mater Chem C Mater* **3**, 10554–10565 (2015).

13. Hodges, J. M. *et al.* Two-Dimensional CsAg5Te3- xSx Semiconductors: Multi-anion Chalcogenides with Dynamic Disorder and Ultralow Thermal Conductivity. *Chemistry of Materials* **30**, 7245–7254 (2018).

14. Vikram, Sahni, B., Jain, A. & Alam, A. Quasi-2D Carrier Transport in KMgBi for Promising Thermoelectric Performance. *ACS Appl Energy Mater* **5**, 9141–9148 (2022).

15. Barreteau, C., Crivello, J. C., Joubert, J. M. & Alleno, E. Looking for new thermoelectric materials among TMX intermetallics using high-throughput calculations. *Comput Mater Sci* **156**, 96–103 (2019).

16. Jain, A. & McGaughey, A. J. H. Thermal transport by phonons and electrons in aluminum, silver, and gold from first principles. *Phys Rev B* **93**, 081206 (2016).

17. Tshitoyan, V. *et al.* Unsupervised word embeddings capture latent knowledge from materials science literature. *Nature* **571**, 95–98 (2019).

18. Pal, K., Park, C. W., Xia, Y., Shen, J. & Wolverton, C. Scale-invariant machine-learning model accelerates the discovery of quaternary chalcogenides with ultralow lattice thermal conductivity. *npj Computational Materials 2022 8:1* **8**, 1–12 (2022).

19. Ren, Z. *et al.* An invertible crystallographic representation for general inverse design of inorganic crystals with targeted properties. *Matter* **5**, 314–335 (2022).

20. Sun, S. *et al.* A data fusion approach to optimize compositional stability of halide perovskites. *Matter* **4**, 1305–1322 (2021).

21. Wagner, J. *et al.* The evolution of Materials Acceleration Platforms: toward the laboratory of the future with AMANDA. *J Mater Sci* **56**, 16422–16446 (2021).

22. MacLeod, B. P. *et al.* Self-driving laboratory for accelerated discovery of thin-film materials. *Sci Adv* **6**, (2020).

23. Burger, B. *et al.* A mobile robotic chemist. *Nature 2020 583:7815* **583**, 237–241 (2020).

24. Kusne, A. G. *et al.* On-the-fly closed-loop materials discovery via Bayesian active learning. *Nature Communications 2020 11:1* **11**, 1–11 (2020).

25. Park, C. W. & Wolverton, C. Developing an improved Crystal Graph Convolutional Neural Network framework for accelerated materials discovery. *Phys Rev Mater* **4**, (2019).

26. Xie, T. & Grossman, J. C. Crystal Graph Convolutional Neural Networks for an Accurate and Interpretable Prediction of Material Properties. *Phys Rev Lett* **120**, 145301 (2018).

27. Wang, A. Y. T., Kauwe, S. K., Murdock, R. J. & Sparks, T. D. Compositionally restricted attention-based network for materials property predictions. *npj Computational Materials 2021 7:1* **7**, 1–10 (2021).

28. Goodall, R. E. A. & Lee, A. A. Predicting materials properties without crystal structure: deep representation learning from stoichiometry. *Nature Communications 2020 11:1* **11**, 1–9 (2020).

29. Askerka, M. *et al.* Learning-in-Templates Enables Accelerated Discovery and Synthesis of New Stable Double Perovskites. *J Am Chem Soc* **141**, 3682–3690 (2019).

30. Gaultois, M. W. *et al.* Data-driven review of thermoelectric materials: Performance and resource onsiderations. *Chemistry of Materials* **25**, 2911–2920 (2013).

31. Ward, L., Agrawal, A., Choudhary, A. & Wolverton, C. A general-purpose machine learning framework for predicting properties of inorganic materials. *npj Computational Materials 2016 2:1* **2**, 1–7 (2016).

32. Jain, A. *et al.* Commentary: The materials project: A materials genome approach to accelerating materials innovation. *APL Materials* vol. 1 011002 Preprint at https://doi.org/10.1063/1.4812323 (2013).

33. Perdew, J. P., Burke, K. & Ernzerhof, M. Generalized gradient approximation made simple. *Phys Rev Lett* **77**, 3865–3868 (1996).

34. Monkhorst, H. J. & Pack, J. D. Special points for Brillouin-zone integrations. *Phys Rev B* **13**, 5188 (1976).

35. Heyd, J. & Scuseria, G. E. Efficient hybrid density functional calculations in solids: Assessment of the Heyd–Scuseria–Ernzerhof screened Coulomb hybrid functional. *J Chem Phys* **121**, 1187 (2004).